\documentclass[final,english]{bullsrsl}[2022/06/15]

\usepackage[latin1]{inputenc}
\usepackage[T1]{fontenc}

\usepackage{natbib} 
\usepackage{graphicx}

\usepackage{amssymb}
\usepackage{multirow} 
\usepackage{newtxtext}
\usepackage[varvw]{newtxmath}
\begin{document}
\title{Study of Wolf-Rayet stars using uGMRT}

\author[affil={1},corresponding]{Anindya}{Saha}
% %\author[affil={2,3}, corresponding]{Hàrry}{Harrisòn}
\author[affil={1}]{Anandmayee} {Tej}
\author[affil={2}]{Santiago} {del Palacio}
% \author[affil={3}]{Mich\"{a}el} {De Becker}
\author[affil={3}]{Micha\"{e}l} {De Becker}
\author[affil={4}]{Paula} {Benaglia}
\author[affil={5}]{Ishwara} {Chandra CH}
\author[affil={6}]{Prachi} {Prajapati}

\affiliation[1]{Indian Institute of Space Science and Technology, Thiruvananthapuram 695 547, Kerala, India}
\affiliation[2]{Department of Space, Earth and Environment, Chalmers University of Technology, Gothenburg, Sweden}
\affiliation[3]{Space sciences, Technologies and Astrophysics Research (STAR) Institute, University of Li\`{e}ge, Belgium}
\affiliation[4]{Instituto Argentino de Radioastronom\'{i}a, CONICET-CICPBA-UNLP, Argentina}
% \affiliation[4]{Instituto Argentino de Radioastronom\'{i}a, CCT-La Plata, CONICET, Argentina}
\affiliation[5]{National Centre for Radio Astrophysics, Pune 411 007, Maharashtra, India}
\affiliation[6]{Physical Research Laboratory (PRL), Navrangpura, Ahmedabad 380 009, Gujarat, India}
\correspondance{anindya.s1130@gmail.com}
\date{5th May 2023}

\maketitle

\begin{abstract}
In recent years, systems involving massive stars with large wind kinetic power have been considered as promising sites for investigating relativistic particle acceleration in low radio frequencies. With this aim, we observed two Wolf-Rayet systems, WR\,114 and WR\,142, using upgraded Giant Meterwave Radio Telescope observations in Band 4 (550--950 MHz) and Band 5 (1050--1450~MHz). None of the targets was detected at these frequencies. 
Based on the non-detection, we report 3$\sigma$ upper limits to the radio flux densities at 735 and 1260 MHz (123 and 66 $\mu$Jy for WR\,114, and 111 and 96 $\mu$Jy for WR\,142, respectively).
The plausible scenarios to interpret this non-detection are presented.   
\end{abstract}

\keywords{stars: Wolf-Rayet, stars: WR\,114, WR\,142, radio continuum: ISM, radiation mechanisms: thermal, non-thermal}

\section{Introduction}
Wolf-Rayet (WR) stars are characterized by powerful supersonic winds with velocities ranging between $\sim 700$--$6000\,\rm km\,s^{-1}$ \citep[e.g.,][]{{Nugis1998},{Hamann2000},{Nugis2000}} and intense mass loss rates $\sim (1$--$5) \times 10^{-5}\,\rm M_{ \odot}\,yr^{-1}$ \citep[e.g.,][]{{Abbott1986},{Leitherer1997},{Chapman1999}}. As these winds propagate, they transfer a large amount of mechanical energy into the surrounding interstellar medium (ISM) and generate strong adiabatic shocks suitable for relativistic particle acceleration via diffusive shock acceleration \citep[DSA;][]{Drury1983}. Observationally, this can manifest in the detection of synchrotron radiation, that is, non-thermal (NT) radio emission produced by relativistic electrons interacting with the local magnetic fields \citep{Blumenthal1970,White1985}. The synchrotron emission, with a spectrum $S_{\nu} \propto \nu^\alpha$ and a canonical spectral index ($\alpha$) of -0.5, is dominant in the low-frequency radio regime.

The energy budget of NT particle acceleration highly depends on the mass loss rate ($\dot{M}$) and the wind velocity ($v_{\rm w}$). It is directly proportional to the wind kinetic power, $P_{\rm kin}$ ($\approx 0.5\,\dot{M}\,v_{\rm w}^2$). Consequently, WR stars have the required conditions to drive efficient particle acceleration. In this paper, we present low-frequency radio observations of two WR stars, WR\,114 and WR\,142, using upgraded Giant Meterwave Radio Telescope (uGMRT). Relevant details of sources are complied in Table~\ref{tab:WRinfo}. With $P_{\rm kin}\,\ge\,10^{38}\,\rm erg\,s^{-1}$, these two stars are promising targets to search for particle acceleration. WR\,114, classified as spectral type WC5, was suggested to be in a binary system with an OB companion \citep{vanderHucht2001}. However, in a recent study, \citet{Sander2012} ruled out the existence of a binary companion based on modelling of optical and ultraviolet spectra using Potsdam WR model atmosphere code. In their \textit{XMM-Newton} observation of WR\,114, \citet{Oskinova2003} did not detect any X-ray emission. WR 142 is an oxygen-rich WR star of the rare spectral type WO2. It is one of only four WO stars identified in Galaxy. X-ray observations using the \textit{XMM-Newton} and \textit{Chandra} telescopes revealed weak but hard X-ray emission with excess absorption below 2 keV \citep{Sokal2010}. These authors suggested NT emission from inverse Compton scattering of stellar photons as a plausible interpretation of the observations.

%----Table---------
\begin{table}
% \begin{center}
\centering
\begin{minipage}{150mm}
\caption{Parameters of WR\,114 and WR\,142. The spectral type is taken from \citet{Smith1968} for WR\,114, and from \citet{Barlow1982} and \citet{Kingsburgh1995} for WR\,142. $D$ is the distance to the source from \textit{Gaia} DR3 data \citep{GaiaDR32022}. $T_{*}$ and $R_{*}$ are the stellar temperature and radius, respectively, $\dot{M}$ is the mass loss rate and $v_{\infty}$ is the terminal wind velocity \citep{Sander2019}. $P_{\rm kin}$ is the wind kinetic power.}
\begin{tabular}{ l c c c c c c  c}
\hline %
  & Spectral type & $D$ & $T_{*}$ & $R_{*}$ & $\dot{M}$ & $v_{\infty}$ & $P_{\rm kin}$ \\
 & &  (kpc) & (kK) & ($\rm R_{\odot}$) & ($\rm M_{ \odot}\,yr^{-1}$) & $\rm(km\,s^{-1})$ & (erg\,$\rm s^{-1}$) \\
 \hline
 WR\,114  & WC5 & 1.97$\pm 0.09$ & 79 & 2.68 & $\rm 3.1\times10^{-5}$ & 3200 & $\rm 1.0\times10^{38}$ \\
 WR\,142  & WO2 & 1.68$\pm 0.04$ & 200 & 0.8 & $\rm 1.6\times10^{-5}$ & 5000 & $\rm 1.3\times10^{38}$ \\
\hline
\end{tabular}
\label{tab:WRinfo}

% \begin{flushleft}
% {\bf Note:} $^{a}$ \citet{Smith1968} for WR\,114, \citet{Barlow1982} and \citet{Kingsburgh1995} for WR\,142; $^{b}$ $D$ is the distance to the source from \textit{Gaia} DR3 data \citep{GaiaDR32022}; $^{c}$ $T_{*}$ and $R_{*}$ are the stellar temperature and radius, respectively, $\dot{M}$ is the mass loss rate and $v_{\infty}$ is the terminal wind velocity \citep{Sander2019}; $^{d}$ $P_{\rm kin}$ is the wind kinetic power.
% % \end{flushleft}
\end{minipage}
% \end{center}
\end{table}
%----Table---------
%%%%%%%%%%%%%%%%%%%%%%%%%%%%%%%%%%%%%%%%%%%%%%%%
\begin{figure*}[!ht]
    \centering
    \includegraphics[width=0.99\linewidth]{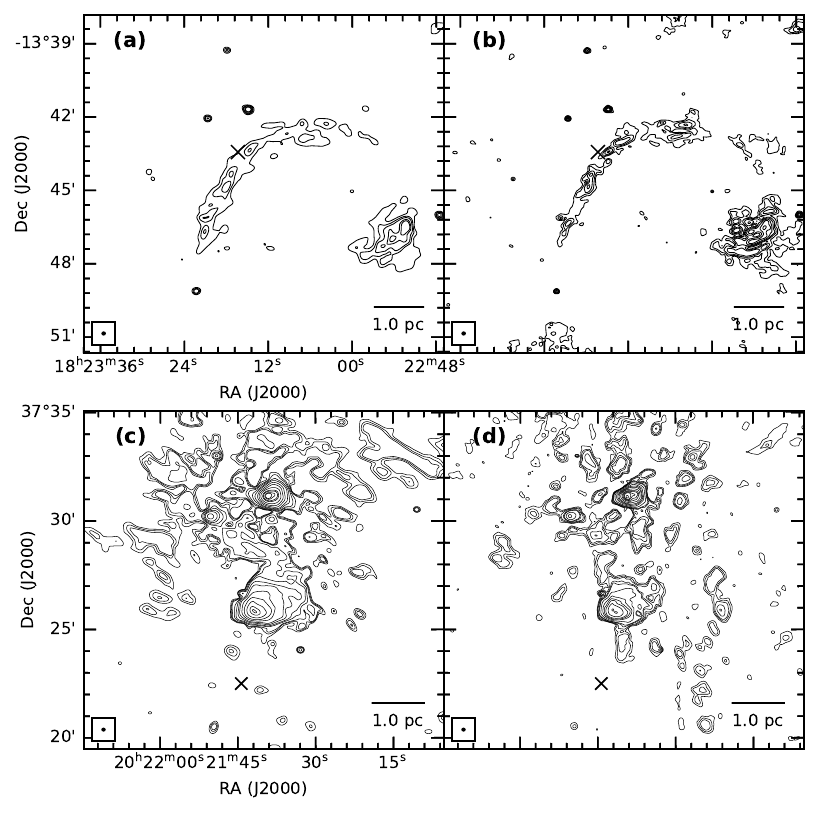}
    \begin{minipage}{12cm}
    \caption{Radio maps of the region around WR\,114 and WR\,142, obtained using uGMRT GWB data. All the maps are convolved to a circular beam of size 5.0$''$. The cross ($\times$) in all panels shows the position of WR star.
    (a) Map of WR\,114 at Band 4 (550--950 MHz). The black contours for the radio emission correspond to the levels of 3, 5, 7, 11, 14 $\times\,\sigma$ ($\rm \sigma = 47\,\mu$Jy\,beam$^{-1}$). 
    (b) Same but for Band 5 (1050--1450 MHz). 
    The contours correspond to the levels of 3, 5, 8, 13, 19, 28, 38 $\times\,\sigma$ ($\rm \sigma = 36\,\mu Jy\,beam^{-1}$).
    (c) Radio map of WR\,142 at Band 4 (550--950 MHz). The black contours correspond to the levels of 3, 5, 7, 15, 25, 50, 90, 200, 400, 800, 1400, 1650, 2000, 2100 $\times\,\sigma$ ($\rm \sigma = 35\,\mu$Jy\,beam$^{-1}$).
    (d) Same but for Band 5 (1050--1450 MHz). 
    The contours correspond to the levels of 3, 5, 9, 17, 23, 55, 120, 190, 280, 500, 880, 1150, 1420, 1600, 1890 $\times\,\sigma$ ($\rm \sigma = 130\,\mu Jy\,beam^{-1}$). In all the panels, the contours are smoothed over 5 pixels using a Gaussian kernel.}  
    \label{fig:radiomaps}
    \end{minipage}
\end{figure*}
%%%%%%%%%%%%%%%%%%%%%%%%%%%%%%%%%%%%%%%%%%%%%%%%%%%
\vspace*{5mm}

\section{Radio continuum emission}

We probed the radio continuum emission associated with WR 114 and WR 142 and their environment in Band 4 (550--950 MHz) and Band 5 (1050--1450 MHz) uGMRT at Pune, India. The details of the GMRT system are described by \citet{Swarup1991} and \citet{Gupta2017}. 3C286 and 3C48 were used as primary flux calibrators. For WR\,114, we selected 1822$-$096 and 1911$-$201 as phase calibrators at Band 4 and Band 5, respectively, and for WR\,142, 2052+365 (for both bands) was observed after each 30~min scan of the target to calibrate the phase and amplitude variations over the entire observing run. The data reduction was done using the
CAsa Pipeline-cum-Toolkit for Upgraded GMRT data REduction (\texttt{CAPTURE}) calibration and imaging pipeline for uGMRT \citep{KaleIshwaraChandra2021}, which uses tasks from Common Astronomy Software Applications \cite[\texttt{CASA};][]{McMullin2007} and Python.
%
%----Table---------
\begin{table*}
\centering
\begin{minipage}{150mm}
\caption{Details of the obtained radio maps.}
\setlength{\tabcolsep}{3.5pt}
\begin{tabular}{ l  c  c  c  c  c}
\hline %
\multirow{2}{*}{} &\multicolumn{2}{c}{WR\,114} & \multicolumn{2}{c}{WR\,142} \\
& Band~4 & Band~5 & Band~4 & Band~5 \\
& 550--950 MHz & 1050--1450 MHz & 550--950 MHz & 1050--1450 MHz\\
\hline
Angular resolution  & 4.3$''\times$3.4$''$ & 2.7$''\times$1.6$''$ & 4.1$''\times$3.3$''$ & 2.3$''\times$1.8$''$\\
\textit{rms} ($\mu$Jy\,beam$^{-1}$) &  41 & 22 & 37 & 32\\
3$\sigma$ upper limits ($\mu$Jy)&  123 & 66 & 111 & 96\\
\hline
\end{tabular}
\label{tab:dataobs}
\end{minipage}
% \end{center}
\end{table*}
%----Table---------
%new para start

\begin{figure}[!ht]
    \centering
    \includegraphics[width=0.5\linewidth]{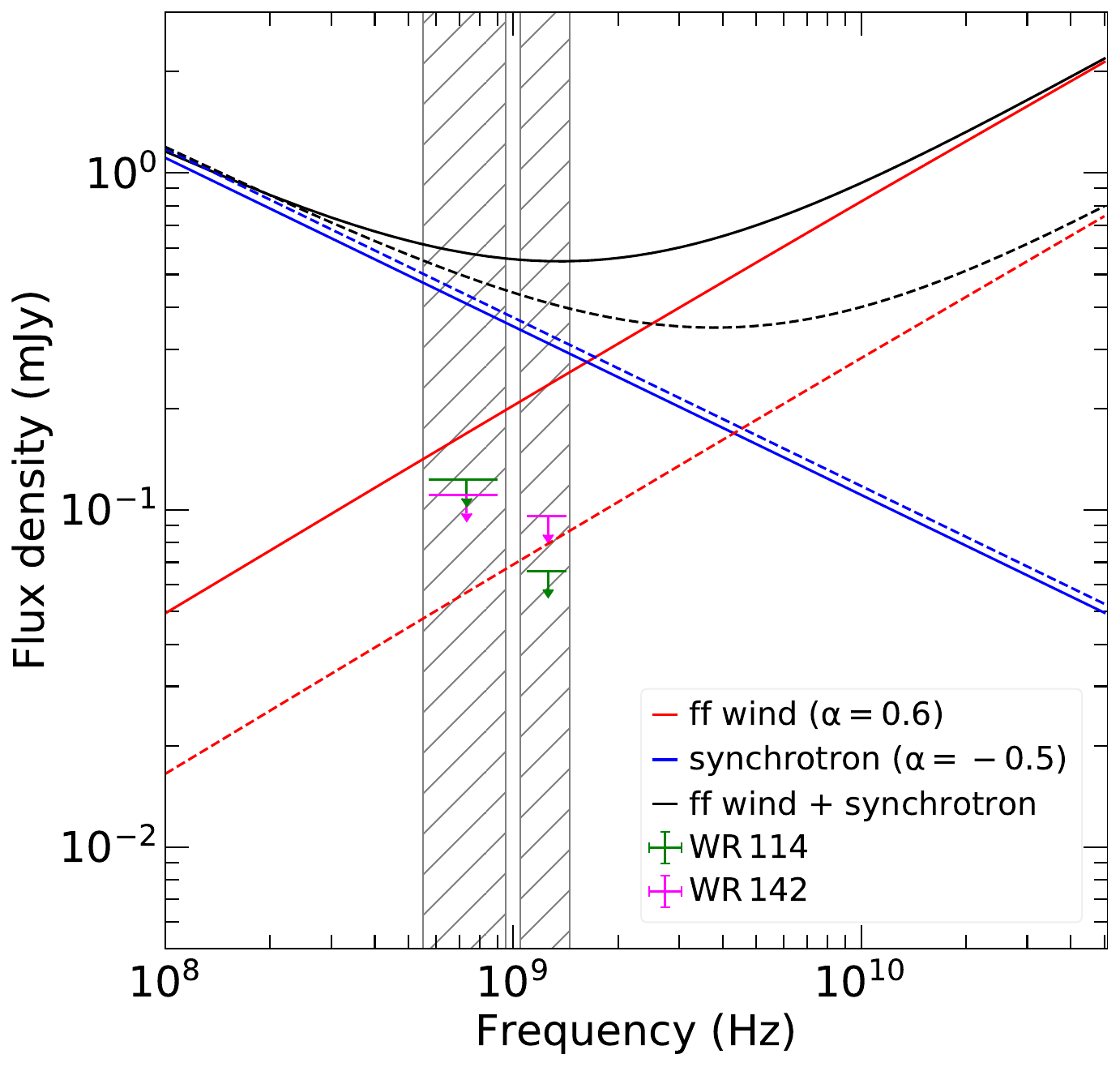}
    \begin{minipage}{12cm}
    \caption{Expected spectral energy distribution of WR\,114 (solid lines) and WR\,142 (dashed lines) at radio frequencies. 
    The red lines represent the estimated free-free emission from the stellar wind, the blue lines represent the lower boundary of the predicted flux density considering NT emission in a colliding-wind binary system, and the black lines are the sum of these two contributions. 
    The 3$\sigma$ upper limits are shown with markers. The shaded regions represent the uGMRT bands used for our observations.}
    \label{fig:sed}
    \end{minipage}
\end{figure}

Figure \ref{fig:radiomaps} presents the uGMRT radio maps of WR\,114 and WR\,142. The map details are compiled in Table \ref{tab:dataobs}. The position of the star is marked in these maps with an `X'. We do not detect radio emission from the stars at either observing frequency. 
The Band 4 map for WR\,114 shows faint, diffuse emission at the location of the WR star. However, this emission is most likely part of the emission from the supernova remnant, SNR G017.4$-$00.1 \citep{{Brogan2006},{Green2009}}.  Based on their empirical relation of radio surface brightness to diameter, \citet{Pavlovic2013} estimated a distance of 18.6~kpc corresponding to the SNR with a large uncertainty (up to 50\%). This distance estimate indicates that the SNR is in the background along the line-of-sight to WR\,114. Thus, these maps provide only upper limits to the flux density from the stars. The $3\sigma$ upper limits derived from the achieved {\it rms} of the uGMRT maps are given in Table~\ref{tab:dataobs} and are plotted in Figure~\ref{fig:sed}.
%%%%%%%%%%%%%%%%%%%%%%%%%%%%%%%%%%%%%%%%%%%%%%%%
%%%%%%%%%%%%%%%%%%%%%%%%%%%%%%%%%%%%%%%%%%%%%%%%%%%

\section{Discussion}
The non-detection of low-frequency radio emission prompts us to explore various scenarios to interpret the observations.
WR stars are often seen to be in binary (or higher multiplicity) systems or are a consequence of binary evolution \citep[e.g.,][]{Meyer2020}. Therefore, in the case of the two WR stars studied here, although there is no conclusive observational evidence of binarity, we cannot rule out this possibility. Hence, in the analysis that follows, we consider two plausible cases: (1) both stars are single stellar systems and (2) they are in colliding-wind binary (CWB) systems.
 
\subsection{Single stellar system}
\label{sec:singlesystem}
In principle, the radio emission from ionised gas in the partially optically thick stellar wind of a single WR star is thermal free-free radiation with a canonical spectral index of 0.6 \citep{{PanagiaFelli1975,Wright1975}}.
For a star with a uniform mass loss rate and isothermal outflow with constant wind velocity, the observed flux density can be written as: 
%%%%%%%%%%%%%%%%%%%%%
\begin{equation}
    \left[\frac{S_{\nu}^{\rm ff}}{\rm Jy} \right] = 23.2\,\left( \left[\frac{\dot{M}}{\rm M_{ \odot}\,yr^{-1}} \right] \left[\frac{\rm km\,s^{-1}}{v_\infty} \right] \frac{ 1}{\sqrt{f_\mathrm{w}}\,\mu}\right)^{4/3} \left( \gamma g_\mathrm{ff}Z^{2}\nu\, \left[\frac{\rm kpc}{D} \right]^3 \right)^{2/3}
    \label{equ:S_nu}
\end{equation}
where the free-free Gaunt factor ($g_\mathrm{ff}$) can be approximated as \citep{Leitherer1991}
\begin{equation}
    g_\mathrm{ff} = 9.77\,\left( 1 + 0.13\log \frac{T^{3/2}_\mathrm{e}}{Z \nu} \right).
   \label{equ:gff}
\end{equation}
%%%%%%%%%%%%%%%%%%%%
%%%%%%%%%%%%%%%%%%%%
Using these expressions, we obtained the predicted flux density values for WR\,114 and WR\,142 by adopting the values of terminal velocity ($v_{\infty}$), mass loss rate ($\dot{M}$), distance ($D$) from Table~\ref{tab:WRinfo}, and considering $T_\mathrm{e} \approx 0.3\,T_{*}$ at radius much greater than $R_{*}$ \citep[see][]{Drew1990}; mean molecular weight, $\mu = 4.0$; {\it rms} ionic charge, $Z = 1.005$; and mean number of electrons per ion, $\gamma = 1.01$ \citep{Leitherer1995}. Further, we considered a volume filling factor of $f_\mathrm{w} = 0.2$ to take into account the stellar wind clumping \citep{Puls2008}.
The predicted radio spectral energy distribution (SED) is shown in Figure~\ref{fig:sed}.
In the case of WR\,114, the uGMRT upper limits are significantly lower than the model-predicted values. A comparison between them indicates that we should have detected radio emission from the star at the $\sim 4\sigma$ and $\sim 10\sigma$ levels in uGMRT Bands 4 and 5, respectively. The non-detection thus implies: (i) the assumed model for the stellar wind is inappropriate for this WR system and/or  (ii) some of the parameters related to the star and/or the wind used in Eqs.~\ref{equ:S_nu} and \ref{equ:gff} to estimate the free-free flux density are inaccurate. 
In the case of WR\,142, the flux densities predicted by the model are consistent with the derived uGMRT upper limits.

\subsubsection{Constraining the mass loss rate in WR\,114}

Since $S_{\nu}^{\rm ff}$ is very sensitive to the mass loss rate (Eq.~\ref{equ:S_nu}), we can use the upper limits on the radio flux density to constrain the mass loss rate of WR\,114. This technique has been widely employed for massive stars in various studies \citep[e.g.,][]{DeBecker2019,Benaglia2019}. \citet{Puls2008} discusses the benefits and limitations of this approach.
Using the flux density upper limit for WR\,114 at Band 5, we estimated a mass loss rate of $\dot{M} \lesssim \rm 1.2\times10^{-5}\,M_{ \odot}\,yr^{-1}$.
A more stringent constraint of $\dot{M} \lesssim \rm 0.7\times10^{-5}\,M_{ \odot}\,yr^{-1}$ is obtained from the probable VLA detection of WR\,114 at 3.6~cm with a flux density of 0.15~mJy reported by \citet{Cappa2004}. In this estimate, we subtracted the contribution to the observed emission from the background SNR in the VLA beam.

\subsubsection{Comparison with WR\,102}
Radio emission associated with single, non-runaway WR stars could also be NT in nature. Theoretically, this could be due to local instabilities \citep[e.g.,][]{Lucy1980,White1985} or magnetic confinement \citep[e.g.,][]{Jardine2001} leading to the acceleration of particles in the stellar winds of single stars. 
% single
Additionally, NT emission can also be seen in the termination shocks of stellar bubbles powered by WR stars. 
These bubbles are formed when powerful stellar winds sweep up the circumstellar material, which could be comprised of the material ejected in previous evolutionary phases or dense ambient interstellar material. 
Such a scenario is revealed for the stellar bubble G2.4+1.4 surrounding the single WR star WR\,102, where synchrotron emission was detected at low uGMRT frequencies \citep{Prajapati2019}. The WR\,102 system is very similar to WR\,142. They are of the same spectral type and have similar wind kinematics ($\dot{M},\,v_\infty$).
While WR\,102 presents a spectacular stellar bubble, there is no such associated bubble detected for WR\,142. The formation of a bubble 
is dependent on several aspects like the progenitor evolutionary phase, wind parameters, mass ejected in previous evolutionary phases, the local ISM density, lifetime in the WR phase, and the proper motion of the star. Comparing the results of WR\,102 and WR\,142 indicates a possible correlation between synchrotron emission and presence of a wind-ISM interaction site where the termination shock in the stellar winds can drive particle acceleration. 
However, more case studies along this line are required to assess the role of the local ISM and its interaction with the powerful WR winds as a necessary ingredient for particle acceleration in single WR stars.

\subsection{Colliding-wind binary systems}
\citet{vanderHucht2001} have suggested the existence of a binary companion to WR\,114 and \citet{Sokal2010} invoked a colliding-wind shock model to explain the higher temperature observed in the X-ray spectra of WR\,142. Hence, based on the above inferences, we investigate both systems in the framework of CWBs.
In the low-frequency radio domain, particle-accelerating colliding-wind binaries (PACWBs) are expected to be composite radio emitters, with a thermal contribution from stellar winds of each binary component which will most likely follow the free-free radiation model discussed in Section~\ref{sec:singlesystem} and a NT contribution of synchrotron emission from the wind-collision region (WCR). 
Following the discussion of energy budget of PACWBs by \citet{DeBecker2013}, it is clear that a small fraction of the total wind kinetic power is converted to the final radio emission after going through a downstream of a series of energy conversion processes. Thus, the radio synchrotron emission is directly proportional to the wind kinetic power, although the proportionality between the two is dependent on several variables (such as wind properties, wind opacity, the particle acceleration efficiency, orbital phase, and the geometry of the WCR) that can differ significantly between systems.

To get an estimate of the synchrotron emission, we calculate the fraction of wind kinetic power that transfers to total radio luminosity ($L_\mathrm{radio}$), which is defined as radio synchrotron efficiency, $RSE = L_{\rm radio}/P_{\rm kin}$ considering the contribution of the thermal component to be negligible. 
By utilizing the available radio measurements and the corresponding $P_{\rm kin}$ for a few selected PACWBs from the catalog \citep{DeBecker2013}, \citet{DeBecker2017} derived empirical equations \citep[see Eqs. 1 and 2 of][]{DeBecker2017} for defining the lower and upper limits of $RSE$. Using these relations, we obtain the $RSE$ ranges from $10^{-9.7}$ to $10^{-7.6}$ and $10^{-9.9}$ to $10^{-7.8}$ for WR\,114 and WR\,142, respectively. 
The $L_\mathrm{radio}$ is estimated from the $RSE$ limits, which can then be converted to flux density ($S_{\nu}^{\rm NT}$) from NT emission at the observing frequency ($\nu$ in Hz) using the following equation:
%%%%%%%%%%%%%%%%%%%%%%%%%
\begin{equation}
    S_{\nu}^{\rm NT} = \frac{10^{26}\,(\alpha + 1)\, {\nu}^{\alpha}\,{L}\rm _{radio}}{4\pi D^2 \,( {\nu}^{(\alpha +1)}_2 - {\nu}^{(\alpha +1)}_1 )} \quad \mathrm{mJy},
\end{equation}
%%%%%%%%%%%%%%%%%%%%%%%%%
where $\alpha$ is the spectral index, and $D$ is the distance (in cm) to the source. Using the lower limit of radio luminosity, $\nu_1 = 0.1$~GHz, $\nu_2 = 50$~GHz, and adopting $\alpha = -0.5$, we estimate the lower limit of the flux density at different frequencies. The resultant SEDs for synchrotron emission are shown in Figure~\ref{fig:sed}.
For both sources, the 3$\sigma$ upper limits from our observations are much lower than the (minimum) expected value. This implies that if NT emission were present, it should have been detected with a high signal-to-noise in our uGMRT observations.

The non-detection of NT emission for WR\,114 and WR\,142 could possibly imply that:
(i) the stars are not in binary systems with a massive companion with strong winds; (ii) the stars are in very wide binary systems (with period of decades), where one of the binary components is far from periastron, resulting in a very low synchrotron luminosity; (iii) the stars are in close binary systems, where radio emission would be drastically reduced due to free-free absorption (FFA) of radio photons by dense stellar wind material.
To further analyse case (iii), we estimate the radius of the wind photosphere ($R_{\nu}$) \citep{{Wright1975}, {Daley2016}} at our observing frequencies. This refers to the distance at which the optical depth equals one. This value sets the lower limit on the size of the region from which radiation at frequency $\nu$ can be observed. 
Considering the same set of parameters as used in Eqs.~\ref{equ:S_nu} and \ref{equ:gff}, we obtain $R_{\nu}$ at the uGMRT frequencies of 735 and 1260~MHz, to be 104 and 71~AU, respectively for WR\,114, and 34 and 23~AU, respectively for WR\,142. Hence, orbital separation would be less than $\sim$70 AU and $\sim$20 AU for WR\,114 and WR\,142, respectively, for FFA to play a significant role and pose a severe obstacle for the emergence of radio photons from the binary components.

\section{Conclusions}

We studied WR\,114 and WR\,142 using uGMRT observations at low frequencies (735 and 1260 MHz). Despite the high sensitivity ($\sim 35\,\mu$Jy\,beam$^{-1}$) and resolution ($\sim 3''$) of our observations, we did not detect any radio emission from either star at either frequency. As a result, we have established 3$\sigma$ upper limits on the radio flux densities of WR\,114 (123 and 66~$\mu$Jy) and WR\,142 (111 and 96~$\mu$Jy) at 735 and 1260 MHz, respectively. To interpret the non-detection we consider two possible scenarios: stellar wind from a single star and a CWB system. In the first framework, we are able to constrain the mass loss rate of WR\,114 to $\dot{M}\,\lesssim$ $\rm 0.7\times10^{-5}\,M_{ \odot}\,yr^{-1}$, which is approximately four times lower than the existing model-based values. Additionally, if we assume both WR stars to be in CWB systems, the lack of NT emission in the uGMRT maps indicates the possibility of either a very wide binary system with orbital period of several decades (not close to periastron) or a close binary system with strong FFA.

\begin{acknowledgments}
This work is supported by the Belgo-Indian Network for Astronomy and astrophysics (BINA), approved by the International Division, Department of Science and Technology (DST, Govt. of India; DST/INT/BELG/P-09/2017) and the Belgian Federal Science Policy Office (BELSPO, Govt. of Belgium; BL/33/IN12). We carried out this work in the framework of the PANTERA-Stars initiative: \url{www.astro.uliege.be/~debecker/pantera}. We thank the GMRT personnel for making these observations possible. GMRT is operated by the National Centre for Radio Astrophysics of the Tata Institute of Fundamental Research.
\end{acknowledgments}

\begin{furtherinformation}

\begin{authorcontributions}
This study was carried out as part of Anindya Saha's PhD research under a long-term collaboration where all co-authors provide contributions.
\end{authorcontributions}

\begin{conflictsofinterest}
The authors declare no conflict of interest.
\end{conflictsofinterest}

\end{furtherinformation}

\bibliographystyle{bullsrsl-en}

\bibliography{reference}

\end{document}